\documentclass[usenatbib]{mn2e}
\usepackage[dvips]{graphicx}

\title[The UDF galaxy luminosity - size relation]{The galaxy luminosity - size relation and selection biases in the Hubble Ultra Deep Field}
\author[E. Cameron and S. P. Driver]{E.~Cameron$^1$\thanks{Visiting fellow at the University of St Andrews} and S.~P.~Driver$^2$\\
$^1$ Research School of Astronomy and Astrophysics, Mount Stromlo Observatory, Cotter Road, Weston Creek, A.C.T., 2611, Australia\\
$^2$ SUPA\thanks{Scottish Universities Physics Alliance}, School of Physics and Astronomy, University of St Andrews, North Haugh, St Andrews, KY16 9SS, Scotland}

\begin{document}
\maketitle

\begin{abstract}
We use the Hubble Ultra Deep Field to study the galaxy luminosity-size ($M$-$R_e$) distribution.  With a careful analysis of selection effects due to both detection completeness and measurement reliability we identify bias-free regions in the $M$-$R_e$ plane for a series of volume-limited samples.  By comparison to a nearby survey also having well defined selection limits, namely the Millennium Galaxy Catalogue,  we present clear evidence for evolution in surface brightness since $z \sim 0.7$.  Specifically, we demonstrate that the mean, rest-frame $B$-band $\left< \mu \right>_{e}$ for galaxies in a sample spanning 8 magnitudes in luminosity between $M_B = -22$ and $-14$ mag increases by $\sim$1.0 mag arcsec$^{-2}$ from $z \sim 0.1$ to $z \sim 0.7$.  We also highlight the importance of considering surface brightness dependent measurement biases in addition to incompleteness biases.  In particular, the increasing, systematic under-estimation of Kron fluxes towards low surface brightnesses may cause diffuse, yet luminous, systems to be mistaken for faint, compact objects.
\end{abstract}


\section{Introduction}
The observational properties of galaxies at any given epoch are a direct result of their formation and evolution histories.  Theories of galaxy formation traditionally adhere to one of two vastly different paradigms---either monolithic collapse (MC) or hierarchical clustering (HC).  In the MC scenario all galaxy types and sizes are created at high redshift by the rapid collapse of primordial gas clouds \citep{egg62,lar75}.  Subsequent evolution is then primarily passive with minimal interaction between nearby neighbours.  Alternatively, under the HC scheme larger galaxies are progressively built from smaller ones during the hierarchical merging of their host dark matter (DM) haloes \citep{whi78}.  Within the HC framework, disc galaxies are the first morphological types formed in the early universe, while ellipticals are constructed from mergers of similar-sized discs over roughly a Hubble time \citep{too72,too77}.\\

Numerous observational studies have provided evidence to support or contradict each of these paradigms, and recent revisions to both theories have been made in light of such findings.  For instance, the homogeneity in the early-type colour-magnitude relation from different clusters \citep{and02} suggests that these galaxies formed at $z > 2$, which favours MC over HC.  However, a number of imaging studies (such as \citealt{bar99}) have reported a deficit of distant ellipticals with passively-evolving colours---contradicting the historic MC scenario of a single burst of star-formation in the early universe.  A so-called `reformed monolithic collapse' model \citep{sch99} has the majority of stars forming at high redshift, but with secondary episodes of star-formation at low redshift caused by internal processes.  \citet{bel04} advocate an adaptation of HC by incorporating `dry mergers', in which the brightest ellipticals grow in size by gas-poor mergers with other ellipticals.  This picture is motivated by their discovery of a factor of 2 increase in stellar mass on the red (i.e., passively evolving) sequence since $z \sim 1$ in the COMBO-17 survey data.  This result is consistent with studies of partially-depleted cores in elliptical galaxies, which indicate such galaxies have experienced, on average, one dry merger \citep{gra04,mer06}. Some authors (such as \citealt{lac85,cay96,bou97}) have developed theories outside of either paradigm using the so-called `backward' approach, whereby high redshift galaxy properties are inferred from detailed studies of the star-formation history of the Milky Way and other local galaxies.  They offer the alternative `infall' model whereby a radially-dependent global star-formation rate means galaxies form from the inside out.  Others, such as \citet{dri06}, advocate in a purely qualitative manner a mixed model in which bulges form first via a rapid collapse or merger phase forming the bulge with subsequent disc formation through splashback and infall.\\

Recent multi-wavelength, deep imaging surveys (e.g.\ the HDFs, GOODS, GEMS, COSMOS and the UDF) have provided a wealth of data for empirical studies of high redshift galaxy evolution, which can test and constrain the above-mentioned formation theories.  For example, \citet{som04} compare the photometric redshift distribution and morphologies of galaxies in the GOODS southern field to theoretical expectations.  In the important $z \ge 1.5$ regime where the models strongly diverge, they observe an excess number density relative to the MC prediction and a deficit relative to the HC one.  However, the disturbed morphologies of objects in their high redshift sample are interpreted as evidence in favour of the general framework of hierarchical formation.  \citet{dad05} present a selection of galaxies at $1.4 \leq z \leq 2.5$ in the UDF with compact, early-type morphologies and spectral energy distributions (SEDs) consistent with passively evolving stellar populations.  They demonstrate that the space density of these galaxies at $\left< z \right> = 1.7$ is only a factor of 2-3 smaller than that of their local counterparts.  At first glance the prevalence of such galaxies in the early universe is difficult to explain within the HC theory, in which luminous ellipticals should be the last galaxy types to form.  However, \citet{del06} argue that `down-sizing' behaviour, for elliptical galaxy star formation, is actually an inherent property of hierarchical formation in their $\Lambda$CDM cosmological simulations.\\

A number of authors have attempted to quantify luminosity and size (or surface brightness) evolution in the galaxy population to high redshift using deep imaging surveys.  For bright galaxies ($M_B \la -18$ mag), the distributions of these two key observables are well constrained locally---both individually, as the luminosity function \citep{nor02,bla03} and size function \citep{she03}, and in bi-variate space, as the luminosity-size distribution (or LSD, \citealt{cro01,she03,dri05}).  It is notoriously difficult to make robust comparisons between galaxy samples drawn from different epochs due to the impact of redshift and surface brightness dependent selection effects, such as the $(1 + z)^4$ cosmological surface brightness dimming.  Fortunately though, the luminosity-size plane is the natural domain in which to confront such observational biases \citep{dis76,phi86,boy95}, and deep, space-based imaging can push back the low surface brightness, faint magnitude and compact size boundaries (see \citealt{dri99}).  \citet{mac05} study early-type, red galaxies in the GEMS survey and find evidence for evolution in the LSD consistent with the passive fading of ancient stellar populations.  In particular, they report a $\sim$1.0 mag increase ($V$-band) at small sizes ($0.5 < R_{50} \leq 1.0$ $h^{-1}$ kpc) to $z = 0.7$ and a $\sim$0.7 mag increase for larger sizes to $z = 1.0$.  Considering disc-dominated galaxies in the GEMS survey, \citet{bar05} find strong evolution of $\sim$1.0 mag arcsec$^{-2}$ in $V$-band surface brightness to $z \sim 1$, but a constant stellar-mass-size relation over this time.  Their results best fit the predictions of the infall model of galaxy formation. \citet{tru05} have recently combined deep, near-IR imaging of the HDF-S and MS1054-03 fields with the SDSS ($z \sim 0.1$) and GEMS ($z \sim 0.2-1$) surveys.  They present evidence of size evolution at fixed luminosity of $(1 + z)^{-0.84 \pm 0.05}$ for early-types and $(1+z)^{-1.01 \pm 0.08}$ for late-types (in rest-frame $V$-band) out to $z \sim 3$, and reach similar conclusions to the other two studies mentioned above.\\

In this paper we use the unprecedented depth of the UDF ACS images \citep{bec06} with the supporting GOODS project NICMOS \citep{tho05} and ISAAC (Vandame et al., in prep.) observations to study the galaxy LSD out to high redshift.  Careful attention is paid to the relevent selection effects to define a bias free region of parameter space in the absolute magnitude-size plane at a range of redshift intervals.  A series of volume-limited samples of UDF galaxies is then defined and the corresponding LSDs constructed.  These are compared to a local benchmark from the Millennium Galaxy Catalogue (MGC) and the degree of evolution quantified.  Finally, we compare our method and results to those of other recent studies in this field.  The outline of this paper is as follows.  Section 2 contains a description of the dataset, as well as the measurement of photometric and structural parameters.  The selection limits are defined in Section 3, and in Section 4 we present evidence of galaxy evolution.  Section 5 presents the comparison of our results to others and in Section 6 we summarise our work and give conclusions.  A cosmological model with $\Omega_{0} = 0.3$, $\Omega_{\Lambda} = 0.7$ and $H_{0} = 100$ km s$^{-1}$ Mpc$^{-1}$ is used throughout.  These specific values of the cosmological parameters were adopted for ease of comparison between the present UDF work and the slightly older MGC results.  Unless otherwise stated, all magnitudes are given in the AB system.\\

\section{The UDF Data}

The Hubble Ultra Deep Field (UDF) consists of an 11 arcmin$^2$ patch of sky centred on RA = $03\degr$ $32'$ $39.0''$, Dec = $-27\degr$ $47'$ $29.1''$ (J2000) in the region of the Fornax Constellation.  The publicly released ACS/WFC Combined Images (version 1.0) span the optical wavelength range 3700 to 10,000 $\AA$ in four wide-band filters : F435W ($B$), F606W ($V$), F775W ($i$) and F850LP ($z$).  The F775W $i$-band image has the longest total exposure time of 347,110 s (144 orbits).  Each single exposure was half an orbit in duration with the pointing cycled through a four part dither pattern.  Each image has been processed through the standard HST data pipeline and drizzled to a pixel scale of 0.03$''$/pixel.  An $i$-band selected catalogue of 10,040 sources (h\_udf\_wfc\_V1\_cat, hereafter referred to as `the on-line catalogue') is also included in the version 1.0 data release---details of its construction are presented in \citet{bec06}.  However, as no Kron magnitudes were provided we elected to generate our own object catalogue as described below.  This also allowed us to make our own decisions regarding deblending of irregular sources.\\

\subsection{Source detection}
A preliminary source extraction was performed using the Starlink implementation (Extractor V1.4-3) of the popular SExtractor package \citep{ber96}.  The threshold for both detection and analysis of our objects was set to a constant surface brightness of 27.395 mag arcsec${}^{-2}$ and a uniform background adopted.  The minimum number of connected pixels to register a detection was set to 9, which is consistent with the size of the PSF FWHM ($\sim$$0.084''$).  Two parameters critical to object deblending are the minimum contrast value and the number of deblending sub-thresholds (see \citealt{ber96}).  Our choices were identical to those used in generating the on-line catalogue, namely 0.03 and 32 respectively.  An abnormally high number of spurious detections were found along the edge of the image mosaic where there are fewer stacked exposures and the signal-to-noise is poor.  Objects with centroids inside these regions, which extend $\sim$100 pixels (3$''$) in from the field boundary, were removed from the source catalogue.  The constant background approach was chosen to avoid additional biases against faint, extended galaxies that can arise in a mesh-based subtraction.  Variation of the mean, local background level over the science-grade $i$-band image was found to be roughly two orders of magnitude smaller than the width of the background noise distribution.  As such it will have a negligible effect on the recovered magnitudes.  The 27.395 mag arcsec${}^{-2}$ level for our detection and analysis threshold was chosen to be similar to that used in extracting the on-line catalogue, which was set to 0.61 times the RMS background noise.  Excluding the low signal-to-noise boundary region decreases the measured width of the background noise.  Thus, although 27.395 mag arcsec${}^{-2}$ corresponds to 0.61 times the full field RMS, it gives an effective limit of 0.91 times the RMS of the interior region actually used for the study.\\

\subsection{Comparison with on-line catalogue}
A sub-sample of 2532 sources with an $i$-band Kron magnitude brighter than 28.0 mag was selected from our preliminary detection list (1.5 mag brighter than the nominal completeness limit derived from the turnover of the counts).  The positions of these objects were compared to those in the on-line catalogue.  A total of 125 had centroids in disagreement by more than 5 pixels (0.15$''$).  These objects, as well as the 50 largest galaxies, were visually inspected.  A pseudo-colour image was generated by combining the $V$-, $i$- and $z$-band WFC/ACS observations.  In each case, our segmentation image was viewed alongside the on-line one, as well as its colour and $i$-band counterparts.  SExtractor appeared to have erroneously deconvolved a single galaxy into multiple sections for 24 of the 50 largest galaxies in our catalogue.  These were restored and 65 redundant sub-components deleted.  Sixty of the 125 objects with mis-matched centroids were also thought to have been poorly deblended.  The most common problem (45 instances) was under-deblending where an apparently close pair of galaxies displayed markedly different colours, suggesting a line-of-sight overlap at different redshifts rather than a single object or merger.  The reverse was true for 9 galaxies with dual nuclei over-deblended.  There were also 6 false detections caused by the diffraction spikes of bright stars.  SExtractor was rerun with four alternative minimum deblend contrast parameter settings (0.001, 0.01, 0.1 and 0.5) to fix these problems.  The final sample of 2497 objects brighter than 28th magnitude will be refered to as iUDF-BRIGHT.  Although the expected 10$\sigma$ limiting magnitude for point sources in the UDF $i$-band image is 29.2 mag \citep{bec06}, we adopt the more conservative 28th magnitude cut-off to ensure a reasonable completeness and reliability in the detection of extended sources (see Section 3.1).  An initial round of star-galaxy separation was performed at this stage using SExtractor's `stellaricity' index---a value between 0 (galaxy) and 1 (star) assigned by an artificial neural network routine for classifying objects.  There is a clear bimodality in the distribution of output stellaricity vs.\ magnitude for iUDF-BRIGHT objects down to 27th mag and we identify 21 certain stars with indices greater than 0.95.  We also confirm another 5 over-exposed stars through a visual inspection of objects brighter than 22nd mag with indices greater than 0.8.  Beyond 27th mag, where there are numerous cases of intermediate stellaricity, we rely upon our photometric redshifts to establish object type.\\

\subsection{Photometric comparison}
\begin{figure}
\includegraphics[width=84mm]{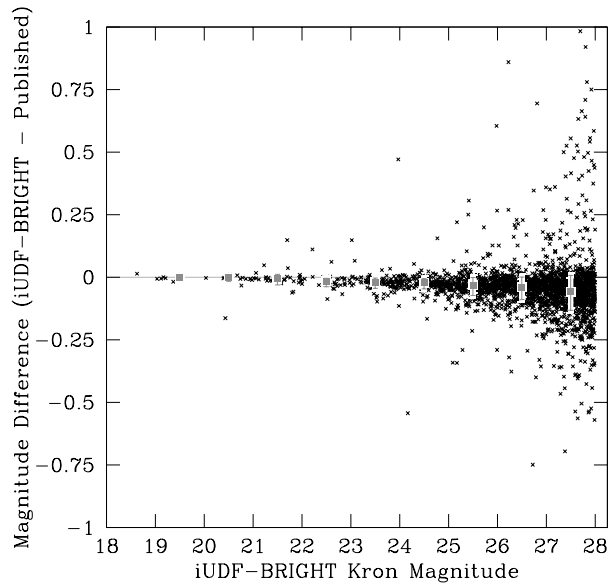}
\caption{\label{pubmatched} Differences between the measured $i$-band isophotal magnitudes of objects in the iUDF-BRIGHT catalogue and those of the on-line one.  This data is plotted as a function of the iUDF-BRIGHT Kron magnitudes used to select our $m <$ 28.0 mag sample.  The mean difference in each magnitude bin is overplotted as a grey square with 1$\sigma$ error bars after 3$\sigma$-clipping.}
\end{figure}

Here we compare our photometry with that of the on-line catalogue.  Fig.\ \ref{pubmatched} contains a plot of the difference between isophotal magnitudes computed for the iUDF-BRIGHT galaxies and the on-line values as a function of our preferred Kron values.  There is a slight difference between the limiting isophote used to compute object flux in each catalogue, and we use a constant background whereas \citet{bec06} use RMS weight maps.  However, with the exception of a number of outliers that were deblended differently, both measurements are generally very similar.  The faintest objects show the largest discrepancy with the 27-28 mag bin having a 3$\sigma$-clipped mean difference of -0.06 mag and a standard deviation of 0.07 mag.  This level of disagreement is not considered significant or problematic given our alternative extraction procedure.\\

\subsection{Half light radii measurements}
The half light radii described in this paper will be defined as the semi-major axis of the elliptical aperture containing half an object's Kron magnitude flux.  This parameter was calculated using a Fortran program that iteratively refines the size of a test elliptical aperture and sums the enclosed light.  The position angle and ellipticity of the test aperture used are those derived by SExtractor during the object detection and analysis phase.  Flux contamination from nearby objects was avoided by excluding pixels attributed to other sources in the segmentation image.  Pixels on the boundary of the test aperture were split into a 10$\times$10 grid in order to estimate their fractional contribution to the enclosed light.  The half light radii of compact objects are affected by the blurring effect of the diffraction limited PSF.  Using the artificial galaxy simulations described in Section 3.2 we find we can recover the true (i.e., intrinsic) size to within 25 per cent accuracy down to $0.05''$ by correcting the measured size according to 
\[\label{hlreqn}
R_{e,\mathrm{intrinsic}}^{2} = R_{e,\mathrm{measured}}^{2} - \alpha \Gamma^{2}
\]
where $\alpha = 0.30$ and $\Gamma = 0.084''$ is the PSF FWHM.  The value of $\alpha$ was chosen to optimise the accuracy of the recovered sizes.  The half light radii derived in this manner were used to calculate the apparent mean effective surface brightness of our objects via the relation
\[
\left< \mu \right>_{e, \mathrm{app}} = m_{\mathrm{Kron}} + 2.5 \log_{10} (2 \pi {R_e}^2 ) \mathrm{.}
\]
This provides a crude inclination correction assuming zero opacity.\\

\subsection{Redshifts}
Photometric redshifts for objects in the iUDF-BRIGHT sample were obtained from two separate sources---from a catalogue supplied by B.\ Mobasher (priv.\ comm., 2005) and from the catalogue of \citet{coe06}, hereafter referred to as M05 and C06 respectively.  By deriving alternative luminosity-size relations using each catalogue in turn and comparing them, we hope to gauge the impact of the potentially large inaccuracies of the photometric redshift approach on our results.  Both sets of redshift estimates were computed using the Bayesian method of \citet{ben00}, but with significant differences in the implementation.  Firstly, M05 uses a fixed 1$''$ aperture to compute object fluxes in each bandpass filter, while C06 use a more sophisticated procedure to derive aperture-matched, PSF-corrected fluxes.  And secondly, although both use the recalibrated SED template library of \citet{ben04}, C06 add two new blue model starburst templates.  These differences lead to large disagreements in the redshifts derived for many of the iUDF-BRIGHT galaxies.  The M05 catalogue consists of 7250 $z$-band selected objects of which 2385 have counterparts in iUDF-BRIGHT (2497 in total), and 73 are identified as having star-like SEDs (leaving 2312 galaxies).  All these galaxies have counterparts in the C06 catalogue.\\  

\begin{figure} 
\includegraphics[width=84mm]{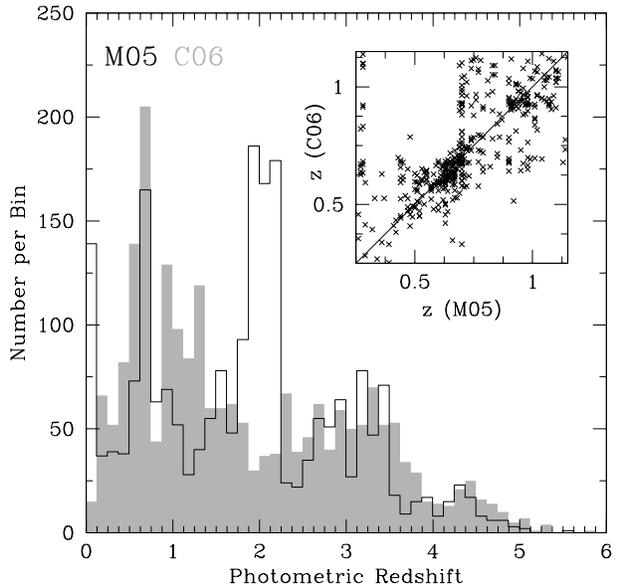}
\caption{\label{z_dist} The distribution of photometric redshifts for galaxies in iUDF-BRIGHT from the M05 (black outline) and C06 (grey shaded) catalogues.  The inset figure is a comparison of individual galaxy redshift estimates in the range relevant to this study, $z =$ 0.25 to 1.15.}
\end{figure}

The iUDF-BRIGHT galaxy redshift distributions from each catalogue are shown in Fig.\ \ref{z_dist} for comparison, and they differ substantially.  M05 finds strong peaks in the bins spanning $z =$ 0-0.125, 0.625-0.75 and 1.875-2.125, whereas C06 find peaks at $z =$ 0.5-0.75, 0.875-1.125 and 1.25-1.375.  It is encouraging, at least, that both detect an overdensity corresponding to the wall identified in the wider CDFS at $z \sim 0.67$ by \citet{lef04} in their spectroscopic survey.  However, \citet{coe06} note that they do not find the $z \sim 0.73$ wall identified in the same survey (while M05 appears to), although the ability of the photometric technique to resolve such close features is questionable.  The disagreement between the two catalogues concerning the redshifts of other regions of overdensity is worrying.  It appears to result mainly from the differences in SED template libraries used since the majority of galaxies in these features are matched in C06 by their bluest starburst templates.  The galaxies in these disputed features are also extremely faint and lack spectroscopic redshifts, so it is impossible to evaluate the merits of each catalogue in this regard.  Hence, we duplicate all analyses using both catalogues in parallel and later investigate the effect of their disagreement on our final results.  We can, however, investigate the accuracy of our photometric redshifts for a small number of bright galaxies with published spectroscopic data.\\

\begin{figure} 
\includegraphics[width=84mm]{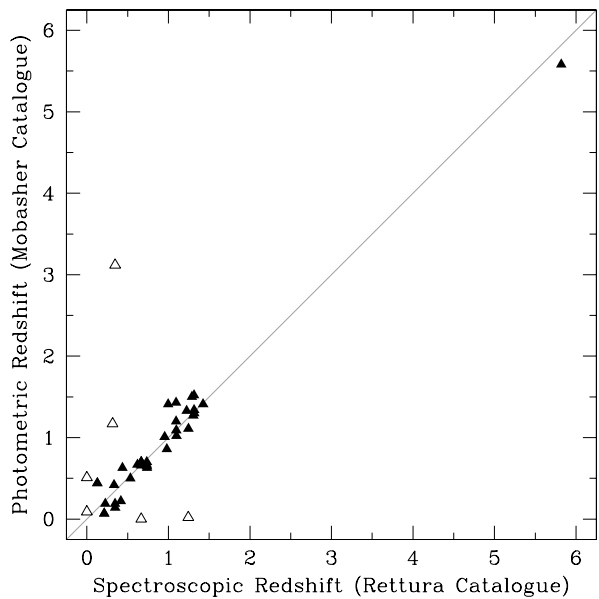}
\caption{\label{comp} A comparison of the M05 photometric redshift estimates with spectroscopically measured values from Rettura's CDF-S Master catalogue (24 matched redshifts from the VIMOS VLT Deep Survey and 18 from VLT FORS2).  With the exclusion of the six outliers (open triangles), redshifts for the remaining 36 galaxies (solid triangles) are all in close agreement with a mean difference, $\Delta z = z_{\mathrm{phot}} - z_{\mathrm{spec}}$, of $0.012 (1+z_{\mathrm{spec}})$ and a standard deviation of just $0.101 (1+z_{\mathrm{spec}})$.  Reasons for the failure of the photometric technique on the six outliers are discussed at the end of Section 2.5.}
\end{figure}

An on-line master catalogue\footnote{http://www.eso.org/science/goods/spectroscopy/CDFS\textunderscore Mastercat/} of published spectroscopic redshifts for objects in the GOODS CDF-S field (encompassing the UDF) is maintained by Rettura.  There are 18 of these from VLT FORS2 observations \citep{van05} with `solid' or `likely' quality flags that match iUDF-BRIGHT objects.  The VIMOS VLT Deep Survey \citep{lef04} provides redshifts with 95 per cent or 100 per cent confidence flags for another 24 members of our sample.  Fig.\ \ref{comp} contains a plot comparing these spectroscopic redshifts to the photometric estimates from the M05 catalogue.  Upon the exclusion of six outliers (from 42), the remaining measurements are in close agreement with a mean difference, $\Delta z = z_{\mathrm{phot}} - z_{\mathrm{spec}}$, of $0.012 (1+z_{\mathrm{spec}})$ and a standard deviation of just $0.101 (1+z_{\mathrm{spec}})$.  In a similar comparison to 41 galaxies in the Rettura catalogue for which they have `reliable' photometric redshift estimates (as identified by their ODDS and $\chi^2_{mod}$ values) C06 find a much smaller standard deviation of only $0.04 (1+z_{\mathrm{spec}})$.  Whilst this comparison certainly provides an indication of the relative accuracy and reliability of the two photometric redshift catalogues for bright galaxies, it cannot be extrapolated to evaluate their performance at faint magnitudes.\\

The six outliers in the M05 comparison serve as case studies of situations in which the photometric redshift technique can break down entirely.  In most instances the problem ultimately stems from incorrect aperture magnitudes.  For example, two spectroscopically confirmed stars ($z_{\mathrm{spec}}=0.000$) were mis-classified as galaxies at $z_{\mathrm{phot}}= 0.090$ and 0.510 respectively because they saturated in one or more of the ACS filters, thereby corrupting their flux measurements.  (We note that these objects would not have been included in our galaxy sample anyway as they were previously identified as stars based on their high SExtractor stellaricity indices.)  Three other outliers had their aperture magnitudes spoilt by contamination from bright, line-of-sight companions.  The remaining outlier ($z_{\mathrm{spec}} = 0.3151$, $z_{\mathrm{phot}} = 1.170$) was for a highly disturbed system with multiple nuclei.  The most likely problem here was that the irregular, young stellar population of this object was poorly represented by any of the standard spectral templates used for the photometric redshift calculations.  The frequency of galaxies with unusually blue, star-forming SEDs in the full iUDF-BRIGHT sample is estimated to be $\sim$6.0 per cent (see Section 2.6 below), which is a relatively minor source of uncertainty in our final results.  However, it does suggest that the use of sophisticated aperture-matched, PSF-corrected fluxes and inclusion of extra blue starburst templates in the C06 method is likely to be beneficial.\\

\subsection{K-corrections}
Individual galaxy K-corrections, from observed $i$-band to rest-frame MGC filter $B$-band \citep{lis03}, were computed as follows using photometric redshifts and broadband magnitudes from both the M05 and C06 catalogues.   ACS/WFC photometry was available for all galaxies in at least two of the four $B$, $V$, $i$ and $z$-bands, plus 49 per cent (1132) and 16 per cent (377) of galaxies had additional NICMOS $J$ and/or $H$-band and ISAAC $K_s$-band photometry respectively.  Total system throughput curves were obtained for the relevant filter plus instrument combinations.  These were then integrated over the redshifted synthetic spectral templates of \citet{pog97} to generate a series of artificial magnitudes in each band.  The library of \citet{pog97} contains 27 model spectra based on three Hubble types (E, Sa and Sc) with stellar population ages in the range 2.2 to 15 Gyr.  An additional flat spectrum was added as a `catch-all' type option for very blue galaxies not adequately represented by the original SED library.  The best fit template for each galaxy was identified via a minimisation of 
\[
\chi^2 = \sum_{m = B,V,i,z,J,H,K} \left[ \frac{(m-m_{\mathrm{artificial}})}{\Delta m} \right]^2
\]
using the quoted errors on the magnitudes in each catalogue.  A total of 221 galaxies (9.6 per cent) returned minimum $\chi^2$ values corresponding to probabilities less than 5 per cent.  These extreme outliers were frequently ($\sim$6.0 per cent of our sample) best-fit by the default flat spectral template, suggesting some level of incompleteness in our synthetic SED library for galaxies caught during their starburst phase.  Once the best fit redshifted template, $f_z (\lambda)$, was identified, the K-correction to MGC $B$-band was computed according to 
\[
K_{i \rightarrow B\mathrm{,MGC}} = 2.5 \log (1+z) + 2.5 \log \left[ \frac{\int_{0}^{\infty} f_z (\lambda) \phi_{B\mathrm{,MGC}} (\lambda) d \lambda}{\int_{0}^{\infty} f_z (\frac{\lambda}{1+z}) \phi_{i} (\lambda) d \lambda} \right]
\]
where $\phi_{i}(\lambda)$ represents the $i$-band filter transmission function and $\phi_{B,MGC}(\lambda)$ that of the MGC $B$-band filter.\\

\begin{figure} 
\hspace{-6mm}
\includegraphics[width=94mm]{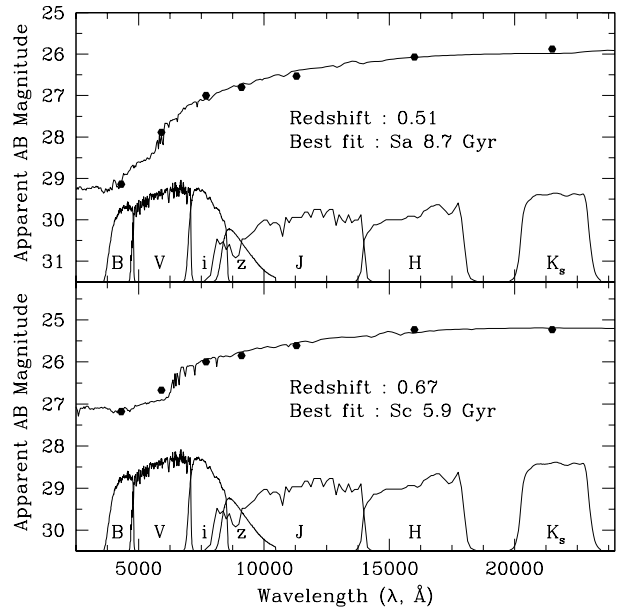}
\vspace{-10mm}
\caption{\label{seds}Best-fit redshifted spectral templates for two example sets of observed band-pass magnitudes (black dots) from the M05 catalogue.  The SED of the first galaxy (top panel) is well described by the Sa type template with $e$-folding time of 7.7 Gyr from the library of \citet{pog97}.  The second galaxy (bottom panel) is well fit by the Sc type template with a 5.9 Gyr $e$-folding time.  The transmission functions of all 7 filters indicate their wavelength coverages.  These have been scaled for clarity and are not intended to represent the total relative throughput in each band.}
\end{figure}

\begin{figure}
\vspace{3mm}
\includegraphics[width=86mm]{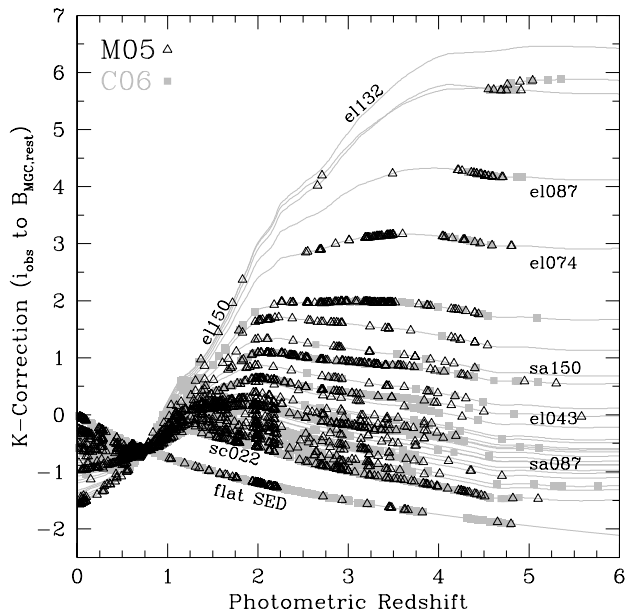}
\vspace{-5mm}
\caption{\label{kcorr}K-corrections from the observed $i$-band magnitude to the rest-frame MGC $B$-band magnitude computed using fluxes and redshifts from both the M05 (black triangles) and C06 (grey squares) catalogues.  The grey lines indicate K-correction functions belonging to each of the 27 spectral templates used from the library of \citet{pog97} (see Section 2.6).  A number of these are labelled according to their Hubble type (`el' = elliptical, `sa' = lenticular  and `sc' = spiral) followed by the age of their model stellar population in units of 0.1 Gyr.}
\end{figure}

Fig.\ \ref{seds} contains examples of the best fit redshifted spectral templates for two galaxies alongside their observed magnitudes in each filter.  Fig.\ \ref{kcorr} contains a plot of all individual K-corrections as a function of redshift, as well as the complete tracks for each model SED.  The reddest K-corrections are for the elliptical type spectra with star-formation timescales of 15 and 13.2 Gyr.  The bluest correction is that for the flat SED, followed by the 2.2 Gyr spiral template.  The difference between our reddest and bluest K-corrections (from observed $i$-band to rest-frame MGC $B$-band) is negligable at $z \sim 0.7$ but grows rapidly thereafter with increasing redshift.  They range $\sim$2 mag by $z=1.5$, $\sim$4.5 mag by $z=2$ and $\sim$8.5 mag by $z=5$.  This has a strong impact on our selection biases at high redshift since the bluest and reddest galaxies of a given $B$-band luminosity will be visible over vastly different volumes.  One way to combat this problem is to use the technique of `band-pass shifting', i.e. to correct the observed magnitude from whichever available filter samples nearest to each galaxy's redshifted, rest-frame $B$-band light.  For objects beyond $z$$\sim$1.5 this would mean correcting from the flux through one of the infrared $J$, $H$ or $K_s$ filters.  Unfortunately, the GOODS and ISAAC observations in these bands are much shallower than the $i$-band image from which our catalogue was selected.  For instance, only 37 per cent of galaxies in our sample have $J$-band photometry and these will be predominantly redder types, which negates the advantages of a smaller K-correction.  Thus, for the present time we will restrict our investigation to redshifts below $\sim$1.5 where the colour bias is less severe.\\

\subsection{Absolute Quantities}
The apparent magnitude ($m$) and mean surface brightness ($\left< \mu \right>_{e, \mathrm{app}}$) of the iUDF-BRIGHT galaxies were converted into absolute quantities using their photometric redshifts and the K-corrections described above.  Luminosity distances ($D_l$) in units of Mpc were calculated according to an $\Omega_{0} = 0.3$, $\Omega_{\Lambda} = 0.7$ and $H_0 = 100$ km s$^{-1}$ Mpc$^{-1}$ cosmological model.  The relevant formulas are 
\[
M = m - 5 \log_{10} (D_l(z)) - 25 - K(z) \mathrm{ ,}
\]
for the absolute magnitude and,
\[
\left< \mu \right>_{e, \mathrm{abs}} = \left< \mu \right>_{e, \mathrm{app}} - 10 \log_{10} (1 + z) - K(z) \mathrm{ ,}
\]
for the absolute mean effective surface brightness.  No evolutionary correction was imposed as this is the unknown we intend to constrain.\\

\section{Selection Effects}
Any imaging survey is restricted and biased in its sampling of the galaxy population by a range of well-documented selection effects \citep*{dis76,dis83,imp97,dri99,cro02,dri05}.  The visibility of a particular galaxy depends on its intrinsic properties (e.g.\ luminosity, scale size, light profile, distance and color) and the nature of the survey imaging data (e.g.\ exposure time, sky brightness, noise, bandpass and seeing).  Furthermore, the accuracy with which a galaxy's true luminosity and scale size may be recovered not only depends strongly on the above mentioned parameters, but also the specific measurement techniques used \citep*{cro04,gra05}.  Understanding the limitations of the iUDF-BRIGHT sample is critical to the robust comparison of absolute magnitude-size distributions at different redshifts.  We have used artificial galaxy simulations to identify a region of minimal selection bias in the apparent magnitude-size plane.  The boundaries of this region are then mapped into the absolute magnitude-size plane via the method of \citet{dri99} for a series of volume-limited samples.\\

\subsection{Apparent Limits---Simulations}
Artificial galaxy simulations are commonly used for estimating survey selection limits (e.g.\ \citealt{tru02,bou04,dri05}).  However, different authors vary significantly in their implementation and interpretation of this technique.  The method used here is as follows.  The luminosity-size plane was divided into a 21x21 grid covering the relevant observational window.  For each grid point the IRAF \texttt{artdata} package was used to generate 100 artificial galaxies with the corresponding size and flux but with random positions in the field and axial ratios (between 0.3 and 1.0).  We use exponential light profiles simulated out to 5 $R_e$ and scaled to account for the flux lost by this truncation.  The simulated galaxy size corresponds to the major axis half light radius.  Each galaxy was convolved with a Gaussian point spread function of 0.084$''$ FWHM.  These objects were inserted into the $i$-band UDF image 25 at a time and SExtractor run to search for them.  The extraction parameters chosen were identical to those used to generate the iUDF-BRIGHT catalogue.  The half light radii of all detected artificial galaxies were measured via the elliptical aperture method.\\

\begin{figure}
\hspace{-2mm}
\includegraphics[width=90mm]{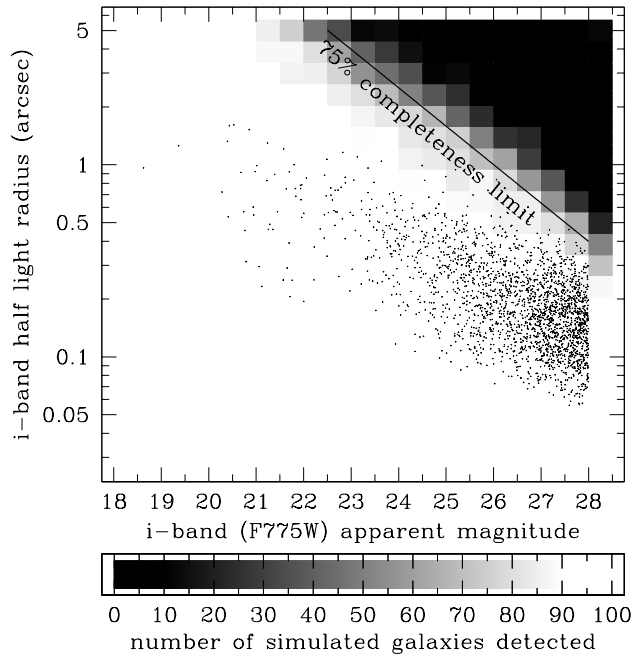}
\caption{\label{completeness_exp} `Completeness' diagram for objects in the UDF $i$-band image computed from our galaxy simulations.  In each bin the grey scale indicates the number of galaxies detected of the 100 inserted with that size and magnitude.  The real, observed iUDF-BRIGHT galaxy population is overlayed as black dots.  The low surface brightness completeness limit, beyond which the detection rate falls below 75 per cent, is also marked.}
\end{figure}

\begin{figure}
\hspace{-2mm}
\includegraphics[width=90mm]{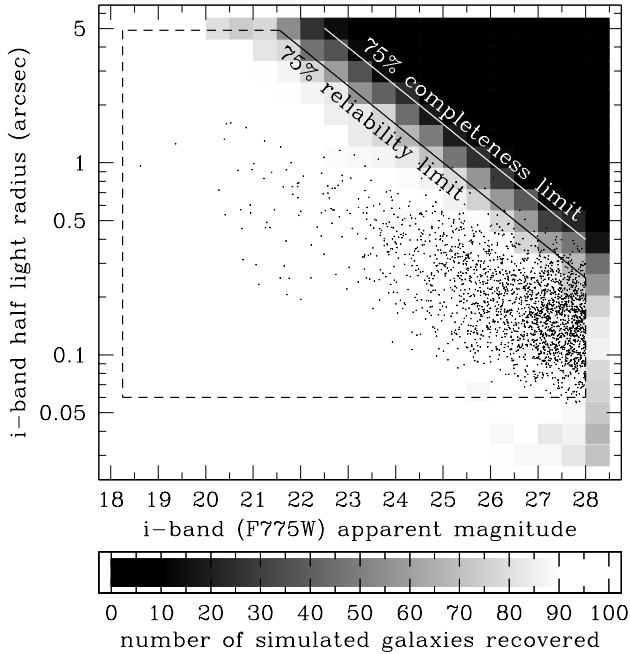}
\caption{\label{reliability_exp}`Recoverability' diagram for objects in the UDF $i$-band image computed from our galaxy simulations.  In each bin the grey scale indicates the number of galaxies detected \textit{with measured fluxes and half light radii within 25 per cent of their input values} of the 100 inserted.  The real, observed iUDF-BRIGHT galaxy population is overlayed as black dots.  The low surface brightness reliability limit, beyond which the recoverability rate falls below 75 per cent, is marked in black too.  The remaining limits making up our five-sided bias free region are illustrated with broken lines.}
\end{figure}

\begin{figure}
\hspace{-2mm}
\includegraphics[width=90mm]{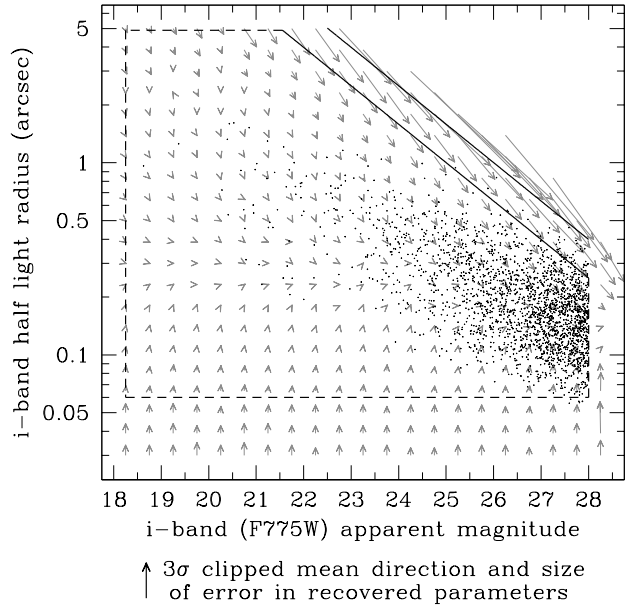}
\vspace{-8mm}
\caption{\label{vectors_exp}`Error vector' diagram for objects in the UDF $i$-band image computed from our galaxy simulations.  The grey arrow eminating from each bin indicates the typical size and direction of the systematic error in our flux and scale size measurements.  Each arrow terminates at the coordinate of the 3$\sigma$-clipped mean magnitude and half light radius of the detected objects from that simulation bin.  The real observed iUDF-BRIGHT galaxy population is overlayed as black dots.  The bias-free region is marked with black lines.}
\end{figure}

Figures \ref{completeness_exp}, \ref{reliability_exp} and \ref{vectors_exp} display the results of our simulations.  In Fig.\ \ref{completeness_exp} we show sample completeness as a function of apparent magnitude and effective radius (i.e., the number of artificial galaxies detected of the 100 inserted at each grid point).  A galaxy is defined as detected if an object is found having a centroid within 5 pixels ($0.15''$) of the input simulation position.  This search radius was made conservatively small to ensure the chance of erroneous matches to existing, real objects was negligable.  In Fig.\ \ref{reliability_exp} we plot `recoverability', which we define to be the number of detected galaxies having measured magnitude and effective radii within 25 per cent of their input values.  And finally, in Fig.\ \ref{vectors_exp} we produce an `error vector' diagram showing the (3$\sigma$-clipped) mean size and direction of the difference between input and recovered values for the detected objects originating in each bin.  Together, the recoverability and error vector diagrams allow one to identify any regions of the observable parameter space contaminated by galaxies with measured magnitudes and half light radii that poorly reflect their true, intrinsic properties.\\

The completeness results plotted in Fig.\ \ref{completeness_exp} indicate that we can detect our simulated galaxies over almost the entire region of the apparent magnitude-size plane spanned by the iUDF-BRIGHT sample.  The only area of low completeness lies in the upper right corner of these plots and corresponds to objects of extremely low mean surface brightness.  In fact, we detect over 75 per cent of simulated objects in our bins out to $\left< \mu \right>_{e} \sim 28.0$ mag arcsec$^{-2}$.  None of the real, observed galaxies are measured to lie within this problem area, and the galaxy population appears to naturally decline in density well before this boundary.  One might be tempted to conclude from this that our sample is not subject to any significant surface brightness dependent selection effects.  However, we have not yet establisted that our flux and scale size measurements are free of bias over the same region of parameter space.\\

The recoverability results in Fig.\ \ref{reliability_exp} reveal that there are biases in the Kron fluxes and half light radii computed for both (apparent, not necessarily intrinsically) faint and low surface brightness galaxies, even in cells of high detection completeness.  These biases were not unexcepted, since they stem from well-documented problems of the Kron magnitude technique \citep{and02,ben04,gra05}.  The Kron magnitudes computed by Sextractor are the sum of light enclosed within an aperture of radius 2.5 times the luminosity-weighted Kron radius (twice the image moment radius).  In theory this should contain 96.0 per cent of the light of a pure exponential profile---a short-fall of just 0.04 mag of the true, instrinsic galaxy flux.  In practice, however, the calculation of the Kron radius is never perfect and is systematically under-estimated in galaxy images with few effective radii sampled above the background sky, i.e., in low surface brightness objects (see \citealt{gra05}, their Section 2.6).  Under-estimating the Kron radius results in an under-estimation of total flux.  This follows on to an under-estimation of the half light radius when the half flux value is derived from the Kron magnitude.  In constructing the recoverability plots of Fig.\ \ref{reliability_exp}, an adjustment was made to allow for the theoretical short-falls in total flux (and, hence, scale size) expected for a correctly measured Kron radius, namely 0.04 mag and 4 per cent of the scale size.  The surplus measurement errors that appear in this plot are then primarily due to mis-calculation of the Kron radius in low surface brightness systems.  The error vector plot in Fig.\ \ref{vectors_exp} illustrates and confirms the expected sense of these errors, which tend to scatter any very low surface brightness galaxies detected towards fainter magnitudes and smaller sizes.  Using these plots we determine the bias-free selection limit $\left< \mu \right>_e = 27.0$ mag arcsec$^{-2}$ below which $\sim$75 per cent of objects have reliably recovered parameters.  This limit is 1 mag arcsec$^{-2}$ brighter than that determined using the simple completeness results.  In addition, we now find it significantly encroaches on the distribution of real, observed galaxies---meaning that the iUDF-BRIGHT sample is clearly not entirely free of surface brightness dependent selection biases.  This illustrates the importance of considering the limits on both completeness and parameter recoverability.\\

The recoverability plots also indicate that cutting our sample at 28th mag was a sensible decision as galaxies brighter than this (and brighter than the surface brightness limit) are recovered at a rate of over 95 per cent in most bins.  Whereas for fainter galaxies in our 28.0-28.5 mag bins the recoverability rate falls to between 75 and 85 per cent due to their very low signal-to-noise in the image.  There is a slight suggestion of a limit on the recoverability of compact objects in this plot, but our simulations are not very realistic in this regard.  In particular, they mask the limitations of our crude PSF modelling.  In the simulations we convolve all our objects with a perfect Gaussian profile of FWHM 0.084$''$ and later correct the measured sizes using Eqn.\ \ref{hlreqn}.  In reality the UDF $i$-band PSF will neither be perfectly Gaussian or of a constant size across the entire image---0.084$''$ was simply the average value computed from our brightest, non-saturated stars, and had a standard deviation of 0.007$''$.  This level of error in our approximation to the true UDF PSF is insignificant for the vast majority of our sample, but would begin to cause problems in objects whose size is similar to that of the PSF (i.e., half light radii $\sim$0.042$''$).  We thus impose a conservative minimum size cut of $r_{\mathrm{min}} = 0.06''$ to exclude such objects and will look to build a more realistic PSF handling scenario into future versions of our simulation procedure.\\

One would expect there to also be a limit on the largest galaxies we could reliably measure in the UDF.  Very extended sources are more likely to overlap with other objects in the line of sight, which leads to deblending difficulties.  Furthermore, galaxies of comparable size to the field of view can cause problems for proper background subtraction.  However, the simulations reveal that neither of these issues prevented the reliable measurement of fluxes and half light radii for the largest profile sizes tested here of 5$''$.  As the largest real object in our sample is 1.8$''$ in half light radius, our observed distribution is clearly not affected by this limit.  But we do illustrate a limit at $r_{\mathrm{max}} = 5''$ in our plots for consistency, and to indicate the direction of it's movement relative to the others at different redshifts.  There is also an effective bright apparent magnitude limit on this survey due to the deliberate choice of a field with no known bright galaxies.  We estimate this from the brightest galaxy in our sample, $m_{\mathrm{bright}} = 18.26$ mag.\\

By considering all the limits described above we define an observational window in apparent space inside which our sample is complete and structural parameters are reliably recovered.  We shall refer to this as the bias-free region.  The full five-sided bias-free region is indicated in Fig.\ \ref{reliability_exp}.\\


\subsection{Absolute Limits}
The selection boundaries derived for the apparent luminosity-size plane are easily extended to the absolute regime for a volume-limited sample via the method of \citet{dri99}.  We define such a sample by binning our data in narrow redshift intervals ($z_{\mathrm{\tiny{low}}}$ to $z_{\mathrm{\tiny{high}}}$), as shown in Fig.\ \ref{mvzplota}.  The constraint on faint absolute magnitudes is given by
\[
M_{\mathrm{faint}} = m_{\mathrm{faint}} - \log_{10}(D_{l}(z_{\mathrm{high}})) -25 - K_{\mathrm{red}}(z_{\mathrm{\tiny{high}}})
\]
where the applicable K-correction is that of the reddest galaxy in our sample.  Likewise, the corresponding constraint on the most luminous galaxies is 
\[
M_{\mathrm{bright}} = m_{\mathrm{bright}} - \log_{10}(D_{l}(z_{\mathrm{low}})) -25 - K_{\mathrm{blue}}(z_{\mathrm{\tiny{low}}})
\]
using the bluest galaxy K-correction.  These limits are illustrated in the magnitude-redshift plot of Fig.\ \ref{mvzplota}, which highlights the problem of working with large K-corrections.  At high redshift the bluest and reddest galaxies in our sample are visible over vastly different volumes, thereby diminishing the luminosity range over which we sample all spectral types evenly.\\

The apparent surface brightness bound on the bias-free region translates to the following absolute surface brightness limit,
\[
\left<\mu\right>_{e, \mathrm{abs,lim}} = \left<\mu\right>_{e, \mathrm{app,lim}} - 10 \log_{10}{(1+z_{\mathrm{high}})} - K_{\mathrm{red}}(z_{\mathrm{high}}) \mathrm{ .}
\]
The dual impact of the $(1+z)^4$ cosmological dimming and the growing red K-correction mean that this is potentially the most restrictive of the selection limits on distant galaxies.  It is shown as a diagonal line on the luminosity-size diagrams in Fig.\ \ref{mvz_cont}.  The maximum and minimum apparent half light radii limits (in arcsec) are converted to absolute scale sizes (in kpc) via the formula
\[
r_{e\mathrm{,abs}}^{\mathrm{max/min}} = \frac{r_{e\mathrm{,app}}^{\mathrm{max/min}}}{3600} \frac{\pi}{180} \frac{D_{l}(z_{\mathrm{low/high}})}{(1 + z_{\mathrm{low/high}})^2} 1000 \mathrm{ .}
\]

\section{Results}
\subsection{Luminosity-Size Diagrams}
\begin{figure}
\hspace{-5mm}
\includegraphics[width=89mm]{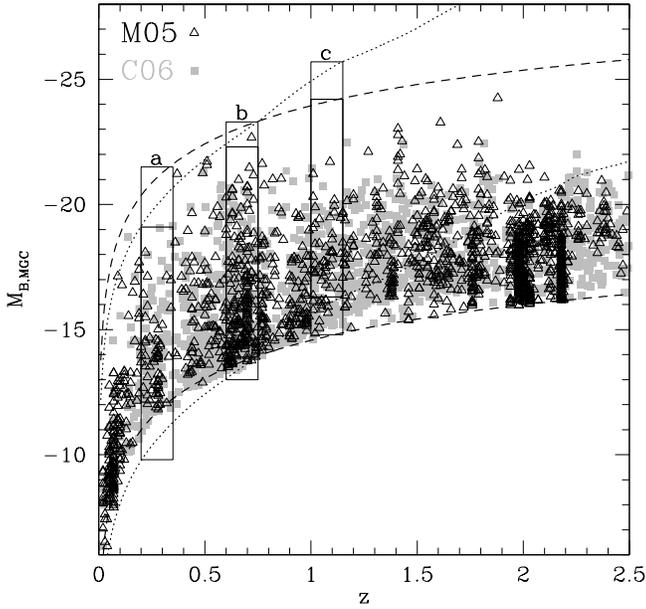}
\vspace{-6mm}
\caption{\label{mvzplota}The absolute magnitude of all iUDF-BRIGHT galaxies with $z < 2.5$ as a function of redshift.  The values derived using the M05 catalogue are shown as black triangles and those using the C06 data as grey squares.  Long and short-dashed lines indicate the upper and lower selection limits in magnitude using K-corrections for the bluest and reddest galaxies respectively.  Volume-limited samples are constructed for three narrow redshift intervals : (a) $z=0.2-0.35$, (b) $z=0.6-0.75$ and (c) $z=1.0-1.15$.  These are designated by the black rectangles; thin lines encompass all objects detected in that redshift range, while the thick lines enclose only those objects within the magnitude selection limits encompassing all spectral types (i.e., for all K-corrections).}
\end{figure}

\begin{figure}
\includegraphics[width=89mm]{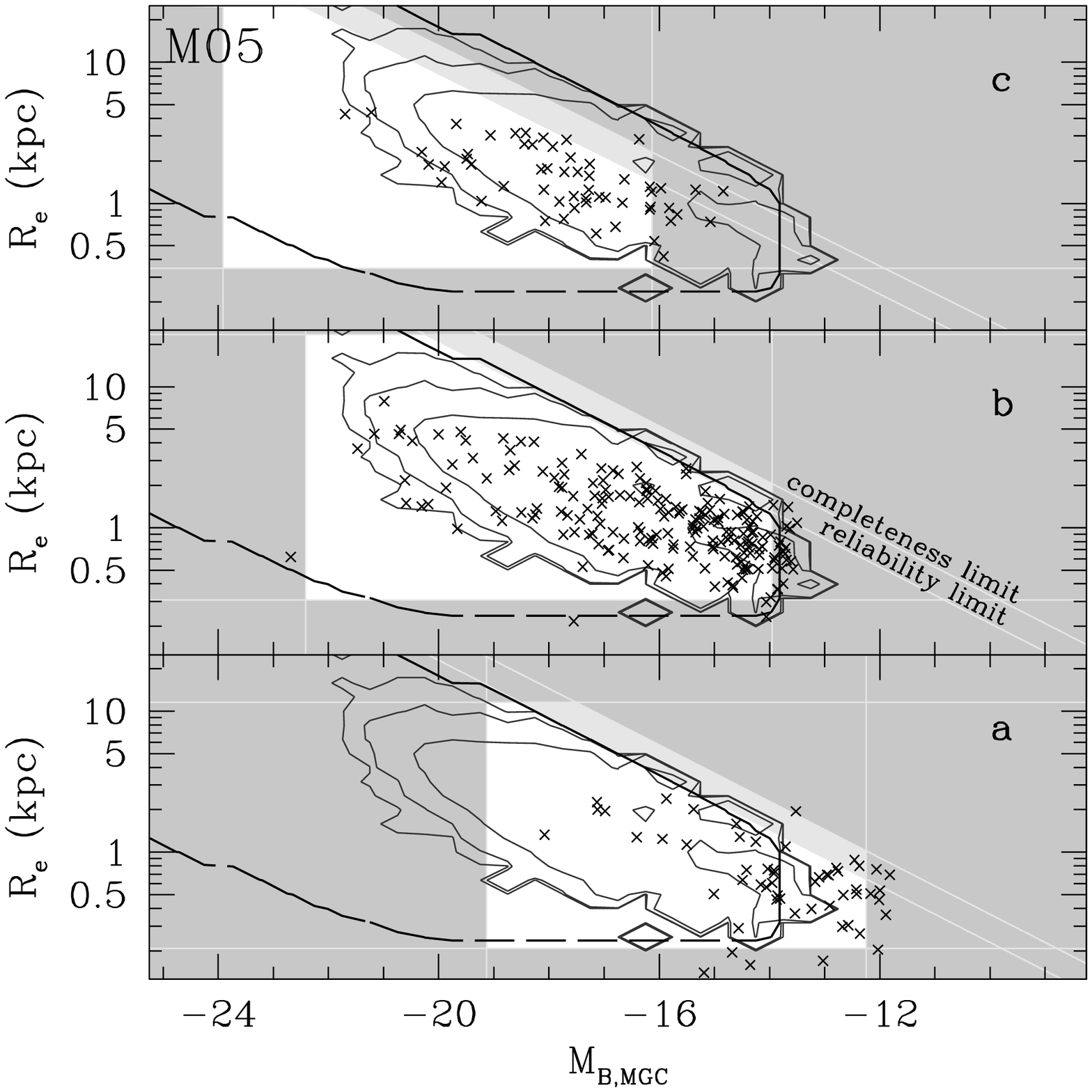}
\includegraphics[width=89mm]{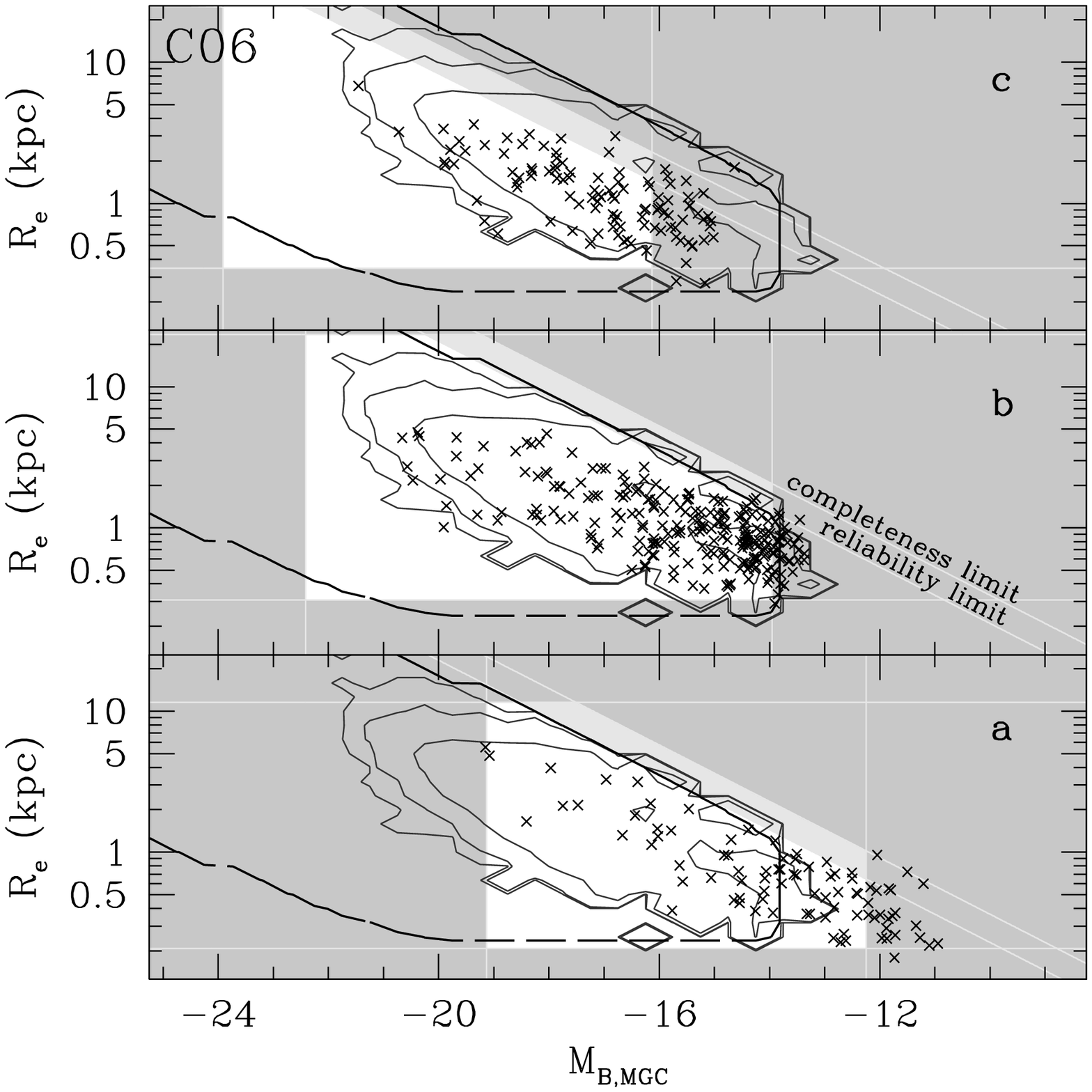}
\vspace{-6mm}
\caption{\label{mvz_cont}Luminosity-size diagrams for each of the three iUDF-BRIGHT volume-limited samples : (a) $z=0.2-0.35$, (b) $z=0.6-0.75$ and (c) $z=1.0-1.15$.  The bias-free selection boundaries (as defined in Section 3.2) are indicated with white lines and grey shading.  The accessable parameter space within is given a white background.  The effect of using the low surface brightness reliability limit in addition to the simple completeness limit is emphasised by plotting both lines and shading the difference in a lighter grey.  Black contours trace the MGC $z=0$ luminosity-size relation at number densities of 10$^{-5 }$, 10$^{-4}$, 10$^{-3}$ and 10$^{-2}$ Mpc$^{-3}$.  The MGC selection boundary as defined by an isovolume contour at 100 Mpc$^3$ is marked with a thick, black, dashed line.  The top diagram is constructed using photometric redshifts from M05 and the bottom using photometric reshifts from C06.}
\end{figure}

In Fig.\ \ref{mvz_cont} we present the $B$-band luminosity-size distribution (LSD) of iUDF-BRIGHT galaxies in three narrow redshift bins: $z = $ (0.2-0.35), (0.6-0.75) and (1.0-1.15) (as shown in Fig.\ \ref{mvzplota}).  These were chosen to lie at, and either side of, the redshift at which the $i$-band filter samples galaxy rest-frame $B$-band light (i.e., where the K-correction is approximately zero).  Selection limits on these volume limited samples were computed as described in Section 3.2 and are overlayed in grey.  Our local ($z = 0.1$) benchmark is derived from the MGC $B$-band bivariate brightness distribution (BBD) in absolute magnitude and surface brightness.  A detailed description of the MGC dataset is contained in \citet{lis03} and construction of the MGC BBD is explained in \citet{dri05}.  The equivalent MGC LSD used here was generated in the same manner, except that the data binning and function fitting were performed in the $L$-$R_e$ plane rather than $L$-$<$$\mu$$>$$^e$.  A contour plot of the resulting number density of MGC galaxies in the $L$-$R_e$ plane is overlain in Fig.\ \ref{mvz_cont} for comparison against our UDF samples at higher redshifts.  To assist in this, the selection boundary of the MGC data (as defined by an isovolume contour at 100 Mpc$^3$) is marked in blue .\\

It is clear from Fig.\ \ref{mvz_cont} that in the interval $z = $ (0.2,0.35) the UDF survey samples a rather different region of the luminosity-size plane than the MGC does locally.  In particular, the extraordinary depth of the UDF imaging and the deliberate pointing away from known bright, nearby galaxies means that at low redshift it primarily detects very faint galaxies (in the range $M = -14$ to $-12$ mag).  The MGC local, bright galaxy sample, on the other hand, is limited to objects with apparent $B$-band magnitudes below 20th mag.  This corresponds to a selection limit of $M_{\mathrm{faint}} < -13.9$ mag at its median redshift ($z=0.1$).  In our highest redshift interval, $z = $ (1.0,1.15), the region of valid comparison on the $L$-$R_e$ plane is also rather small.  At these high redshifts the bias-free window of parameter space accessible with the iUDF-BRIGHT sample only covers a fraction of the local MGC relation in the bright, compact regime.  It is in the intermediate redshift sample at $z = $ (0.6,0.75) that the UDF and MGC observational windows best coincide.  Here we sample the full width of the $z \sim 0.7$ LSD over almost 9 mags with only a slight bias against low surface brightness galaxies for our faintest objects at $M_B \sim -16$ to $-14$ mag.  According to the M05 photometric redshifts, there are 169 iUDF-BRIGHT galaxies in this volume-limited sample that lie within the selection boundaries of both surveys (and 212 for the C06 catalogue).  An eyeball comparison suggests the UDF objects have a similar distribution to the MGC ones, but are somewhat brighter and more compact, suggesting moderate evolution in the LSD to these redshifts.  We quantify this via a 2D K-S test below.\\

\subsection{2-D Kolmogorov-Smirnov Test}
The Kolmogorov-Smirnov (K-S) test provides an estimate of the probability that two distributions are drawn from the same population.  In the one-dimensional K-S test this probability is computed from the maximum cumulative difference between the two distributions.  In its extension to two dimensions the integrated probability in each of four quadrants around a given point forms the basis of the evaluation \citep{pre92}.  The implementation used here is \texttt{ks2d2s} from the Numerical Recipes library, which is valid for sample sizes greater than 20 objects.  Its output probability estimate is less accurate above values of 20 per cent, although probabilities greater than this do correctly indicate that the two distributions being compared are very similar.  Our primary input sample for this test consists of the 169 iUDF-BRIGHT galaxies (M05 redshift catalogue) within the interval $z = $ (0.6,0.75) and contained inside both the UDF and MGC selection limits.  The first comparison we make is to a set of 1000 galaxies drawn from the MGC $z \sim 0.1$ $L$-$R_e$ BBD using a basic Monte Carlo technique.  The K-S test result is a probability of 19 per cent that these two samples are drawn from the same population, which is only a mild degree of similarity by this measure.\\

\begin{figure}
\centering
\includegraphics[width=84mm]{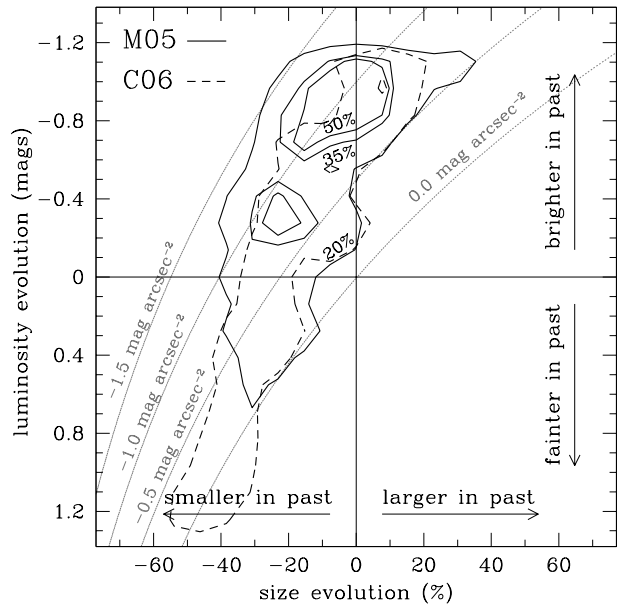}
\caption{\label{kstesta} Results of the 2-D K-S test for different scalings of the MGC $z \sim 0.1$ luminosity-size distribution compared to the iUDF-BRIGHT $z =$(0.6,0.75) sample.  The contours show the output probability value, which is indicative of the likelihood that the higher $z$ UDF LSD could have been drawn from a population described by the scaled MGC LSD.  The two 50 per cent contours enclose our best fit scenarios peaking at $\Delta L  = -0.9$ mag, $\Delta R_e = 5$ per cent and $\Delta L = 0.3$ mag, $\Delta R_e = -25$ per cent.  To aid the reader, lines of constant mean surface brightness evolution are marked in grey, spanning null evolution to the case of galaxies being 1.5 mag arcsec$^{-2}$ brighter in the past.}
\end{figure}

In order to constrain evolutionary scenarios, we examine whether scaling the MGC LSD in luminosity and/or scale size can produce a higher K-S test probability than the case of null evolution.  Our method was to generate a mock MGC data set of 1000 galaxies for each trial MGC LSD scaling and then run the 2-D K-S test to compare it to the UDF sample.  We do this for a broad range of scenarios from galaxies being 1.3 mag fainter to 1.3 mag brighter, and from 70 per cent smaller to 70 per cent larger.  The resulting probability values are displayed as a contour plot in Fig.\ \ref{kstesta}.  It is clear from this figure that there is a wide range of scalings providing a higher degree of similarity to the $z \sim 0.7$ UDF LSD than the $z \sim 0.1$ MGC LSD with no evolution.  The best fits are found in two separate regions of this parameter space.  The first corresponds to mainly luminosity evolution with galaxies being typically $\sim$$0.7$-$1.1$ mag brighter at $z = 0.675$ than they are at $z \sim 0.1$, and between $\sim$$20$ per cent smaller and 10 per cent larger.  The peak of this region is at $\Delta L = -0.9$ mag and $\Delta R_e = -5$ per cent.  The second region of good fit corresponds to galaxies being on average $\sim$$0.3$ mag brighter and $\sim$$25$ per cent smaller in the past.  These two likely evolutionary scenarios both equate to similar degrees of surface brightness evolution, $\sim$1.0 mag for the first and $\sim$0.9 mag for the second, which is necessary to bring the ridge lines of the two LSDs into agreement.  The first case with greater luminosity evolution offers a superior fit to the bright end of the distribution than the second one.  Since the bright end of the UDF $z = 0.675$ LSD has the most reliably measured magnitudes and scale sizes and is well clear of our selection limits, one should attach greater importance to the fit there.  As the K-S test does not allow for such a weighting to be set explicitly, we simply note that $\Delta L = -0.9$ mag with $\Delta R_e = -5$ per cent is our preferred result, but that we cannot rule out the case of $\Delta L = -0.3$ mag with $\Delta R_e = -25$ per cent.\\

Repeating this analysis with the iUDF-BRIGHT LSD derived from the C06 photometric redshift catalogue we find a broad agreement with the evolution predicted using the M05 redshifts.  In particular, the 20\% contours of each K-S test are very similar and isolate essentially the same region of parameter space.  The only difference is that the C06 results also allow the possibility of galaxies having been substantially fainter and smaller in the past (by up to 1.2 mag and -60\%).  This scenario arises because of the large number of faint blue galaxies found at $z \sim 0.6$$-$0.75 in the C06 analysis.  This over-abundance of faint objects relative to the local MGC LSD also means that none of our simple scaled evolutionary scenarios provide $>$35\% probabilities in the K-S test.  Until more reliable (i.e., spectroscopic) redshifts are available for a significant number of these faint systems it will be impossible to properly characterise the faint galaxy population at these redshifts.  For now we can only acknowledge the difficulties we face in this type of study and make the best use of the data available to us.  It is difficult to estimate a formal uncertainty on our most likely evolutionary fit because of the inaccuracy of the K-S test above 20$\%$ probability.  However, as our errors are overwhelmingly dominated by those in the photometric redshift estimates we simply acknowledge the full range of acceptable fits for both catalogues (as shown in Fig.\ \ref{kstesta}).  This includes surface brightness evolution to $z \sim 0.7$ spanning a 0.5 mag arcsec$^{-2}$ dimming to a 1.5 mag arcsec$^{-2}$ brightening.\\

\section{Discussion}
A number of previous studies have explored evolution of the galaxy LSD beyond the local universe using deep imaging data.  This field of investigation came of age with the launch of the Hubble Space Telescope, which allowed the first high resolution, optical imaging studies of the high redshift galaxy population (e.g.\ \citealt{drw95,dri95}).  In one of the earliest such studies \citet{sch95} analysed HST $B$- and $I$-band images of 32 galaxies at $0.5<z<1.2$ randomly selected from the Canada-France Redshift Survey (CFRS).  They identified objects in their sample as being similar to the local mix of morphological types, based on an eyeball classification, with the exception of 9 galaxies dominated by blue compact components.  After also performing bulge-disc decomposition, they found 15 normal late type galaxies with $B/T < 0.5$ and disc luminosities $M_{B}^{\mathrm{AB}} - 5 \log{h_{50}} < -19.8$ mag.  Compared to the local Freeman value, the $B$-band central surface brightness of these objects was higher by 1.2 mag arcsec$^{-2}$, which \citet{sch95} attribute to evolution of the disc galaxy luminosity function.  \citet{roc98} obtained a sample of 270 galaxies by combining spectroscopic redshifts from various sources, including the CFRS, with HST imaging ($B$, $I$ and in some cases $V$ and $U$-band) from the HDF and other surveys.  They investigated the LSD at a range of $z$ intervals and found $B$-band surface brightness evolution of $0.95 \pm 0.22$ mag between $z \sim 0.2$ and $z \sim 0.9$, with similar evolution for all morphological types.  The authors explain this with an `inside-out' style disc formation model, whereby the half light radius increases with time.  However, the results of these studies have been questioned by later works in which selection effects have been given a more in-depth consideration, such as \citet{sim99} and \citet{rav04}.\\

\citet{sim99} studied the LSD of a sample of 190 field galaxies in the Groth Survey Strip with HST $V$ and $I$-band imaging and spectroscopic redshifts from the DEEP survey.  Bulge-disc decomposition was performed to identify disc-dominated systems ($B/T < 0.5$) and extract structural properties.  The disc LSD was then constructed at a series of redshift intervals and surface brightness measured to evolve by 1.3 mag in the rest-frame $B$-band from $z \sim 0.2$ to 0.8, a similar amount to earlier studies such as \citet{sch95} and \citet{roc98}.  The authors then recalculated these LSDs with a weighting that essentially applied the selection function of the highest redshift bin to that of the lower redshift bins.  This was done in order to account for observational incompleteness biases against faint and low surface brightness objects, and to ensure the comparison was being made over the same range of luminosities at all redshifts.  The result was that no detectable mean surface brightness evolution was observed over the redshift range 0.1 to 1.1 for discs with $-19 < M_{B}^{\mathrm{Vega}} - 5 \log{h_{70}} < -22$ mag.  This conclusion is supported by \citet{rav04} in a study using the HST GOODS images and photometric redshifts.  They also restrict their analysis to the galaxies in the lower redshift bins that fall within the selection boundaries on their highest redshift bin at $z=1.0-1.25$.  The disc size function for objects within these bounds is found to remain constant over the range $0.25 < z < 1.25$.\\

The strict selection function approach has recently been criticised for making inadequate use of the information provided by the observed galaxy LSD at each redshift.  In the latest studies, such as \citet{bou04} and \citet{bar05}, the authors instead attempt to establish whether or not the surface brightness (or size) distribution at each redshift is biased by incompleteness effects.  \citet{bar05} combined HST imaging from GEMS with COMBO-17 photometric redshifts to search for evolution in the disc galaxy LSD and stellar mass-size relations out to $z \sim 1$.  They identified disc-dominated systems by their global Sersic index ($n < 2.5$) and used artificial galaxy simulations to estimate their completeness function in the apparent magnitude-size plane.  The faint magnitude limit restricts them to the study of objects with $M_V < -20$ at $z \sim 1.0$, and they applied this limit to their sample at all redshifts to avoid biases in mean surface brightness due to the slope of the LSD.  At each redshift interval they then constructed histograms of the surface brightness distribution, weighting by the completeness function, and estalished that each was approximately Gaussian.  Under the assumption that the disc galaxy surface brightness distribution is intrinsically, roughly Gaussian and uni-modal at all redshifts, they then went on to conclude that they were not missing significant numbers of galaxies at any redshift.  Finally, they fit a linear relation to the mean surface brightness as a function of redshift and found a slope of $-1.43 \pm 0.07$ in rest-frame $B$-band, i.e., an increase of $0.96$ mag arcsec$^{-2}$ to $z=0.67$.\\

\citet{bou04} used a slightly different approach to establish whether their high redshift galaxy samples are affected by low surface brightness incompleteness problems.  They constructed a UBVi-dropout sample set at redshifts $z \sim 2.5-6.0$ from the HDF, GOODS and UDF images.  For each filter dropout sample they compared the GOODS galaxy apparent magnitude-size distribution with that from the deeper UDF (and UDF-P) imaging.  As the primary effect of pushing back the surface brightness completeness limits via the additional UDF exposure time was to add compact objects at the faint magnitude limit, the authors concluded that the shallower GOODS data is essentially complete at bright magnitudes ($-19.7 < M_{1700} < -21.07$).  From the galaxies in this luminosity range they measured size evolution of $(1+z)^{-1.05 \pm 0.21}$ to $z \sim 6$.  \citet{tru05} found a similar degree of evolution in their study, which used J, H and K-band imaging from the VLT FIRES data to probe $z > 1$ optical sizes.  After dividing their sample into bulge and disc-dominated system by global Sersic index (cut at $n=2.5$), the authors compared the observed size distribution of high redshift objects to completeness limits derived from simulations.  As the number of observed galaxies decreases more rapidly towards larger sizes than do the completeness limits, they argue that incompleteness is not biasing the data.  Relative to the SDSS luminosity-size relation they find galaxies with $L_{V} > 3.4 \times 10^{10} h_{70}^{-2} L_{\sun}$ at $z \sim 2.5$ are $\sim$3.5 times smaller than for equally luminous galaxies today.\\

Here in our study of the UDF we have used yet another approach to analysing selection biases and quantifying evolution in the galaxy LSD.  Firstly, we have performed detailed artificial galaxy simulations to establish both the completeness limits and reliability limits of our data, noting particularly the increasing systematic under-estimation of total flux by Kron magnitudes towards low surface brightnesses.  And secondly, by using a local galaxy survey also with well defined selection boundaries (the MGC) we are able to identify a broad region of the LSD over which both samples are free of bias.  This enabled us to establish clear evidence of evolution in mean surface brightness of the galaxy population for a range of over 8 mag in luminosity.  Specifically, we found an increase of 1.0 mag arcsec$^{-2}$ from $z \sim 0.1$ to $z = 0.675$ for objects with $-22 < M_B < -14$ mag.  Assuming a linear trend with redshift we can extrapolate this to a 1.05 mag arcsec$^{-2}$ increase from $z=0$ to $z \sim 0.7$.  This is in agreement with the 0.96 mag arcsec$^{-2}$ to $z \sim 0.67$ recently found by \citet{bar05}, and indeed with the earlier results of \citet{sch95} and \citet{roc98}.  It is also consistent with the surface brightness evolution of $\sim$1.2 mag arcsec$^{-2}$ predicted by the \citet{bou04} fit to $z > 2.5$ size evolution.  In summary, we confirm the evolution in mean surface brightness of the bright galaxy population observed in these previous studies, and demonstrate that it holds down to $M_B \sim -14$ mag.\\

It should be noted that our comparison to the \citet{bar05} and \citet{sch95} results is imperfect because they specifically measured evolution of disc dominated systems, whereas we examine the galaxy population as a whole.  As the UDF and MGC are both field galaxy surveys one would expect a majority of late-type systems.  In fact, eyeball classification of the MGC sample within the selection boundaries defined by \citet{dri05} found $\sim$34 per cent early-type systems.  Our decision not to attempt a morphological subdivision of our sample meant we were able to avoid the added complications of choosing the most meaningful and robust criteria on which to classify galaxies at vastly different redshifts and from different datasets.  However, this is somewhat of a limitation on our ability to use these results to distinguish between galaxy formation scenarios, which generally offer predictions for the evolution of specific morphological classes.  In a future paper we will present a detailed structural analysis of the 169 objects in our $z=0.675$ volume-limited sample, quantify evolution by galaxy type, and compare our findings to theoretical expectations.\\

\section{Summary}
We constructed the iUDF-BRIGHT sample from all galaxies detected in the UDF ACS $i$-band image brighter than 28.0 mag.  Half light radii were measured for these objects and photometric redshifts matched to each from the Mobasher catalogue.  Individual K-corrections were computed using SED template fits to their observed $B$, $V$, $i$, $z$, $J$, $H$ and $K$-band fluxes.  This allowed the derivation of rest-frame, $B$-band absolute magnitudes and scale sizes using a $\Omega_{0} = 0.3$, $\Omega_{\Lambda} = 0.7$ , $H_{0} = 100$ km s$^{-1}$ Mpc$^{-1}$
cosmological model.  Detailed artificial galaxy simulations were then used to establish detection completeness and measurement reliability limits in the observational apparent magnitude-size plane.  We mapped these into the luminosity-size plane and presented the UDF galaxy LSD for a series of volume limited samples.  By comparison to the LSD of the Millennium Galaxy Catalogue, a nearby galaxy survey with well defined selection limits, we identified a region of the $M$-$R_e$ plane over which the $z \sim 0.1$ and $z \sim 0.7$ galaxy populations can be compared free of selection biases.  Evolution was quantified via a 2D K-S test and an increase of $\sim$1.0 mag found for the average surface brightness of galaxies with luminosities $M_B = -22$ to $-14$ mag.  This is in agreement with the results of other recent studies, such as \citet{bou04}, \citet{bar05} and \citet{tru05}, but contradicts the null evolution findings of \citet{sim99} and \citet{rav04}.\\

An important result to emerge from our artificial galaxy simulations was that surface brightness dependent measurement errors are a significant source of potential bias in the observed LSD.  We found that, although exponential profile galaxies were detectable with a completeness of 75 per cent down to $\sim$28 mag arcsec$^{-2}$ in the UDF $i$-band image, flux and scale sizes could only be recovered to within 25 per cent accuracy down to $\sim$27 mag arcsec$^{-2}$ (based on SExtractor Kron magnitudes and sizes).  As our error vector diagram indicates (see Fig.\ \ref{vectors_exp}), the impact of these measurement errors is to \textit{recover extended, low surface brightness galaxies as faint, compact objects}.  If this effect is not accounted for the observed LSD will be be doubly biased---with an under representation of large, diffuse systems and an over abundance of small ones.  This problem of recoverability is only likely to get worse towards higher $z$ and must now be included in all analyses.\\

\section*{Acknowledgements}
We would like to thank Bahram Mobasher of the Space Telescope Science Institute for kindly making available his photometric redshift catalogue for this work.  We also thank Alister Graham of Mount Stromlo Observatory for helpful critcism and advice, and proof reading of this paper.\\


\begin{thebibliography}{}
\bibitem[\protect\citeauthoryear{Aguerri \& Trujillo}{2002}]{tru02} Aguerri J. A. L., Trujillo I., 2002, MNRAS, 333, 633-641
\bibitem[\protect\citeauthoryear{Andreon}{2002}]{and02} Andreon S., 2002, AAP, 382, 495-502
\bibitem[\protect\citeauthoryear{Barden et al.}{2005}]{bar05} Barden M., Rix H.-W., Somerville R. S., Bell E. F., Haubler B., Peng C. Y., Borch A., Beckwith S. V. W., Caldwell J. A. R., Heymans C., Jahnke K., Jogee S., McIntosh D. H., Klaus M., Sanchez S. F., Wisotzki L., Wolf C., 2005, ApJ, 635, 959-981
\bibitem[\protect\citeauthoryear{Barger et al.}{1999}]{bar99} Barger A. J., Cowie L. L., Trentham N., Fulton E., Hu E. M., Soongaila A., Hall D., 1999, ApJ, 117, 102-110
\bibitem[\protect\citeauthoryear{Beckwith et al.}{2006}]{bec06} Beckwith S. V. W., Stiavelli M., Koekemoer A. M., Caldwell J. A. R., Ferguson H. C., Hook R., Lucas R. A., Bergeron L. E., Corbin M., Jogee S., Panagia N., Robberto M., Royle P., Somerville R. S., Sosey M., 2006, astro-ph/0607632
\bibitem[\protect\citeauthoryear{Bell et al.}{2004}]{bel04} Bell E. F., Wolf C., Klaus M., Rix H.-W., Borch A., Dye S., Kleinheinrich M., Wisotzki L., McIntosh D. H., 2004, ApJ, 608, 752-767
\bibitem[\protect\citeauthoryear{Benitez}{2000}]{ben00} Benitez N., 2000, ApJ, 536, 571-583
\bibitem[\protect\citeauthoryear{Benitez et al.}{2004}]{ben04} Benitez N., Ford H., Bouwens R., Menanteau F., Blakeslee J., Gronwall C., Illingworth G., Meurer G., Broadhurst T. J., Clampin M., \& 26 coauthors, 2004, ApJs, 150, 1-18
\bibitem[\protect\citeauthoryear{Bertin \& Arnouts}{1996}]{ber96} Bertin E., Arnouts S., 1996, AAPs, 117, 393-404
\bibitem[\protect\citeauthoryear{Blanton et al.}{2003}]{bla03} Blanton M. R., Hogg D. W., Bahcall N. A., Neta A., Baldry I. K., Brinkmann J., Csabai I.; Eisenstein D., Fukugita M., Gunn J. E., Ivezic Z., \& 11 coauthors, 2003, ApJ, 592, 819
\bibitem[\protect\citeauthoryear{Bouwens et al.}{1997}]{bou97} Bouwens R. J., Cayon L., Silk J., 1997, ApJ, 489, L21-L24
\bibitem[\protect\citeauthoryear{Bouwens et al.}{2004}]{bou04} Bouwens R. J., Illingworth G. D., Blakeslee J. P., Broadhurst T. J., Franx M., 2004, ApJ, 611, L1-L4
\bibitem[\protect\citeauthoryear{Boyce \& Phillipps}{1995}]{boy95} Boyce P. J., Phillipps S., 1995, AAP, 296, 26 
\bibitem[\protect\citeauthoryear{Cayon et al.}{1996}]{cay96} Cayon L., Silk J., Charlot S., 1996, ApJ, 467, L53-L56
\bibitem[\protect\citeauthoryear{Chiosi \& Carraro}{2002}]{chi02} Chiosi C., Carraro G., 2002, MNRAS, 335, 335-357
\bibitem[\protect\citeauthoryear{Coe et al.}{2006}]{coe06} Coe D., Benitez N., Sanchez S. F., Jee M., Bouwens R., Ford H., 2006, AA, 132, 926-959
\bibitem[\protect\citeauthoryear{Cole et al.}{2000}]{col00} Cole S., Lacey C. G., Baugh C. M., Frenk C. S., 2000, MNRAS, 319, 168-204
\bibitem[\protect\citeauthoryear{Cross \& Driver}{2002}]{cro02} Cross N., Driver S. P., 2002, MNRAS, 329, 579-587
\bibitem[\protect\citeauthoryear{Cross et al.}{2001}]{cro01} Cross N., Driver S. P., Couch W. C., Baugh C. M., Bland-Hawthorn J., Bridges T., Cannon R., Cole S., Colless M., Collins C., \& 22 coauthors, 2001, MNRAS, 324, 825-841
\bibitem[\protect\citeauthoryear{Cross et al.}{2004}]{cro04} Cross N. J. G., Driver S. P., Liske J., Lemon D. J., Peacock J. A., Cole S., Noberg P., Sutherland W. J., 2004, MNRAS, 349, 576-594
\bibitem[\protect\citeauthoryear{Daddi et al.}{2005}]{dad05} Daddi E., Renzini A., Pirzkal N., Cimatti A., Malhotra S., Stiavelli M., Xu C., Pasquali A., Rhoads J. E., Brusa M., di Serego Alighieri S., Ferguson H. C., Koekemoer A. M., Moustakas L. A., Panagia N., Windhorst R. A., 2005, ApJ, 626, 680-697
\bibitem[\protect\citeauthoryear{Dalcanton, Spergel \& Summers}{Dalcanton et al.}{1997}]{dal97} Dalcanton J. J., Spergel D. N., Summers F. J., 1997, ApJ, 482, 659-676
\bibitem[\protect\citeauthoryear{de Jong \& Lacey}{2000}]{dej00} de Jong R. S., Lacey C., 2000, ApJ, 545, 781-797
\bibitem[\protect\citeauthoryear{De Lucia et al.}{2006}]{del06} De Lucia G., Springel V., White S. D. M., Croton D., Kauffmann G., 2006, MNRAS, 366, 499-509
\bibitem[\protect\citeauthoryear{Disney}{1976}]{dis76} Disney M. J., 1976, Nature, 263, 573-575
\bibitem[\protect\citeauthoryear{Disney \& Phillipps}{1983}]{dis83} Disney M., Phillipps S., 1983, MNRAS, 205, 1253-1265
\bibitem[\protect\citeauthoryear{Driver}{1999}]{dri99} Driver S. P., 1999, ApJ, 526, L69-L72
\bibitem[\protect\citeauthoryear{Driver, Windhorst \& Griffiths}{1995}]{drw95} Driver S. P., Windhorst R. A., Griffiths R. E., 1995, ApJ, 453, 48
\bibitem[\protect\citeauthoryear{Driver et al.}{1995}]{dri95} Driver S. P., Winidhorst R. A., Ostrander E. J., Keel W. C., Griffiths R. E., Ratnatunga K. U., 1995, ApJ, 499, L23
\bibitem[\protect\citeauthoryear{Driver et al.}{2005}]{dri05} Driver S. P., Liske J., Cross N. J. G., De Propris R., Allen P. D., 2005, MNRAS, 360, 81
\bibitem[\protect\citeauthoryear{Driver et al.}{2006}]{dri06} Driver S. P., Allen P. D., Graham A. W., Cameron E., Liske J., Ellis S. C., Cross N. J. G., De Propris R., Phillipps S., Couch W. J., 2006, MNRAS, 368, 414-434 
\bibitem[\protect\citeauthoryear{Efstathiou et al.}{1998}]{efs98} Efstathiou G., Ellis R. S., Peterson, B. A., 1988, MNRAS, 232, 431
\bibitem[\protect\citeauthoryear{Eggen, Lynden-Bell \& Sandage}{Eggen et al.}{1962}]{egg62} Eggen O. J., Lynden-Bell D., Sandage A. R., 1962, ApJ, 136, 748
\bibitem[\protect\citeauthoryear{Fall \& Efstathiou}{1980}]{fal80} Fall S. M., Efstathiou G., 1980, MNRAS, 193, 189-206
\bibitem[\protect\citeauthoryear{Freeman}{1970}]{fre70} Freeman K. C., 1970, ApJ, 160, 811-830
\bibitem[\protect\citeauthoryear{Graham}{2004}]{gra04} Graham A. W., 2004, ApJ, 613, L33-L36
\bibitem[\protect\citeauthoryear{Graham \& Driver}{2005}]{gra05} Graham A. W., Driver S. P., 2005, PASA, 22, 118-127
\bibitem[\protect\citeauthoryear{Impey \& Bothun}{1997}]{imp97} Impey C., Bothun G., 1997, ARA\&A, 35, 267-307
\bibitem[\protect\citeauthoryear{Kormendy}{1977}]{kor77} Kormendy J., 1977, ApJ, 217, 406
\bibitem[\protect\citeauthoryear{Lacey \& Fall}{1985}]{lac85} Lacey C. G., Fall S. M., 1985, ApJ, 290, 154
\bibitem[\protect\citeauthoryear{Larson}{1975}]{lar75} Larson R. B., 1975, MNRAS, 173, 671-699
\bibitem[\protect\citeauthoryear{Le Fevre et al.}{2004}]{lef04} Le Fevre O., Vettolani G., Paltani S., Tresse L., Zamorani G., Le Brun V., Moreau C., Bottini D., Maccagni D., Picat J. P., \& 40 coauthors, 2004, A\&A, 428, 1043 
\bibitem[\protect\citeauthoryear{Lilly et al.}{1995}]{lil95} Lilly S. J., Tresse L., Hammer F., Crampton D., Le Fevre O., 1995, ApJ, 455, 108
\bibitem[\protect\citeauthoryear{Lilly et al.}{1998}]{lil98} Lilly S., Schade D., Ellis R., le Fevre O., Brinchmann J., Tresse L., Abraham R., Hammer F., Crampton D., Colless M., \& 3 coauthors, 1998, ApJ, 500, 75-94 
\bibitem[\protect\citeauthoryear{Liske et al.}{2003}]{lis03} Liske J., Lemon D. J., Driver S. P., Cross N. J. G., Couch W. J., 2003, MNRAS, 344, 307L
\bibitem[\protect\citeauthoryear{Liske et al.}{2006}]{lis06} Liske J., Driver S. P., Allen P. D., Cross N. J. G., de Propris R., 2006, MNRAS, 369, 1547-1565
\bibitem[\protect\citeauthoryear{McIntosh et al.}{2005}]{mac05} McIntosh D. H., Bell E. F., Rix H.-W., Wolf C., Heymans C., Peng C. Y., Somerville R. S., Barden M., Beckwith S. V. W., Borch A., \& 7 coauthors, 2005, ApJ, 632, 191-209
\bibitem[\protect\citeauthoryear{Mao, Mo \& White}{Mao et al.}{1998}]{mao98} Mao S., Mo H. J., White S. D. M., 1999, MNRAS, 297, L71-L75
\bibitem[\protect\citeauthoryear{Merrit}{2006}]{mer06} Merritt D., 2006, astro-ph/0603439
\bibitem[\protect\citeauthoryear{Mo, Mao \& White}{Mo et al.}{1998}]{mo98} Mo H. J., Mao S., White S. D. M., 1998, MNRAS, 295, 319-336
\bibitem[\protect\citeauthoryear{Mobasher et al.}{2004}]{mob04} Mobasher B., Idzi R., Benitez N., Cimatti A., Cristiani S., Daddi E., Dahlen T., Dickinson M., Erben T., Ferguson H. C., \& 12 coauthors, 2004, ApJ, 600, L167-L170
\bibitem[\protect\citeauthoryear{Nipoti et al.}{2003}]{nip03} Nipoti C., Londrillo P., Ciotti L., 2003, MNRAS, 342, 501-512
\bibitem[\protect\citeauthoryear{Norberg et al.}{2002}]{nor02} Norberg P., Cole S., Baugh C. M., Frenk C. S., Baldry I., Bland-Hawthorn J., Bridges T., Cannon R., Colless M., Collins C., and 18 coauthors, 2002, MNRAS, 336, 907
\bibitem[\protect\citeauthoryear{Peebles}{1969}]{pee69} Peebles P. J. E., 1969, ApJ, 155, 393
\bibitem[\protect\citeauthoryear{Phillipps \& Disney}{1986}]{phi86} Phillipps S., Disney M. J., 1986, MNRAS, 221, 1039-1048 
\bibitem[\protect\citeauthoryear{Phillips, Davies \& Disney}{Phillips et al.}{1990}]{phi90} Phillips S., Davies J. I., Disney M. J., 1990, MNRAS, 242, 235-240
\bibitem[\protect\citeauthoryear{Poggianti}{1997}]{pog97} Poggianti B. M., 1997, A\&A Supp., 122, 399-407
\bibitem[\protect\citeauthoryear{Press}{1992}]{pre92} Press W. H., Teukolsky S. A., Vettering W. T., Flannery B. P., 1992, Numerical recipes in Fortran. The art of scientific computing., 2nd edn., Camberidge University Press, New York
\bibitem[\protect\citeauthoryear{Ravindranath et al.}{2004}]{rav04} Ravindranath S., Ferguson H. C., Conselice C., Giavalisco M., Dickinson M., Chatzichristou E., de Mello D., Fall S. M., Gardner J. P., Grogin M. A., Hornschemeier A., Jogee S., Koekemoer A., Kretchmer C., Livio M., Mobasher B., Somerville R., 2004, ApJ, 604, L9-L12
\bibitem[\protect\citeauthoryear{Roche et al.}{1998}]{roc98} Roche N., Ratnatunga K., Griffiths R. E., Im M., Naim A., 1998, MNRAS, 293, 157-176
\bibitem[\protect\citeauthoryear{Schade et al.}{1995}]{sch95} Schade D., Lilly S. J., Crampton D., Hammer F., le Fevre O., Tresse L., 1995, ApJ, 451, L1-L4
\bibitem[\protect\citeauthoryear{Schade et al.}{1999}]{sch99} Schade D., Lilly S. J., Crampton D., Ellis R. S., Le Fevre O., Hammer F., Brinchmann J., Abraham R., Colless M., Glazebrook K., Tresse L., Broadhurst T., 1999, ApJ, 525, 31-46
\bibitem[\protect\citeauthoryear{Shen et al.}{2003}]{she03} Shen S., Mo H. J., White S. D. M., Blanton M. R., Kauffmann G., Voges W., Brinkmann J., Csabai I., 2003, MNRAS, 342, 978-994
\bibitem[\protect\citeauthoryear{Simard et al.}{1999}]{sim99} Simard L., Koo D. C., Faber S. M., Sarajedini V. L., Vicki L., Vogt N. P., Phillips A. C., Gebhardt K., Illingworth G. D., Wu K. L., 1999, ApJ, 519, 563-579
\bibitem[\protect\citeauthoryear{Somerville et al.}{2004}]{som04} Somerville R. S., Moustakas L. A., Mobasher B., Gardner J. P., Cimatti A., Conselice C., Daddi E., Dahlen T., Dickinson M., Eisenardt P., Lotz J., Papovich C., Renzini A., Stern, D., 2004, ApJ, 600, L135-L138
\bibitem[\protect\citeauthoryear{Thompson et al.}{2005}]{tho05} Thompson R. I., Illingworth G., Bouwens R., Dickinson M., Eisenstein D., Fan X., Franx M., Riess A., Rieke M. J., Schneider G., Stobie E., Toft S., Van Dokkum P., 2005, ApJ, 130, 1, 1-12
\bibitem[\protect\citeauthoryear{Toomre}{1977}]{too77} Toomre A., 1977, in Tinsley B. M., Larson R. B., eds, The Evolution of Galaxies and Stellar Populations., New Haven, Yale University Observatory, p.401
\bibitem[\protect\citeauthoryear{Toomre \& Toomre}{1972}]{too72} Toomre A., Toomre J., 1972, ApJ, 178, 623
\bibitem[\protect\citeauthoryear{Trujillo \& Aguerri}{2004}]{tru04} Trujillo I., Aguerri J. A. L., 2004, MNRAS, 355, 82-96
\bibitem[\protect\citeauthoryear{Trujillo et al.}{2005}]{tru05} Trujillo I., Forster Schreiber N. M., Rudnick G., Barden M., Franx M., Rix H-W., \& 11 others, 2005, astro-ph/0504225
\bibitem[\protect\citeauthoryear{Vanzella et al.}{2005}]{van05} Vanzella E., Cristiani S., Dickinson M., Kuntschner H., Moustakas L. A., Nonino M., Rosati P., Stern D., Cesarsky C., Ettori S., Ferguson H. C., Fosbury R. A. E., Giavalisco M., Haase J., Renzini A., Rettura A., Serra P., \& the GOODS Team, 2005, A\&A, 434, 53-65
\bibitem[\protect\citeauthoryear{White \& Rees}{1978}]{whi78} White S. D. M., Rees M. J., 1978, MNRAS, 183, 241-358
\bibitem[\protect\citeauthoryear{Wolf et al.}{2003}]{wol03} Wolf C., Meisenheimer K., Rix H.-W., Borch A., Dye S., Kleinheinrich M., 2003, A\&A, 401, 73
\end{thebibliography}
\end{document}